\documentclass[11pt,a4paper]{article}
\usepackage{jheppub}
\usepackage[latin9]{inputenc}
\usepackage{amsmath}
\usepackage{amssymb}
\usepackage{graphicx}
\usepackage{babel}
\usepackage[caption=false]{subfig}
\usepackage{mathtools}
\usepackage{slashed}

\newcommand{\e}{\mathrm{e}}

\author[a]{Patrick~Copinger}
\emailAdd{copinger@hiroshima-u.ac.jp}
\affiliation[a]{International Institute for Sustainability with Knotted Chiral Meta Matter (WPI-SKCM$^2$), Hiroshima University, 1-3-1 Kagamiyama, Higashi-Hiroshima, Hiroshima 739-8531, Japan}
\author[b,c,a]{Minoru~Eto}
\emailAdd{meto@sci.kj.yamagata-u.ac.jp}
\affiliation[b]{Department of Physics, Yamagata University, 
Kojirakawa-machi 1-4-12, Yamagata, Yamagata 990-8560, Japan}
\affiliation[c]{Research and Education Center for Natural Sciences, Keio University, 4-1-1 Hiyoshi, Yokohama, Kanagawa 223-8521, Japan}
\author[d,c,a]{Muneto~Nitta}
\emailAdd{mune.nitta@gmail.com}
\affiliation[d]{Department of Physics, Keio University, 4-1-1 Hiyoshi, Yokohama, Kanagawa 223-8521, Japan}

\title{Fermi gas of domain-wall Skyrmions in QCD in a strong magnetic field}
\abstract{
Fermionic domain-wall Skyrmions arise in the chiral soliton lattice of two flavor chiral perturbation theory in a magnetic field 
as the ground state 
and are associated with baryons of the underlying theory of quantum chromodynamics. We analyze the electromagnetic screening characteristics of domain-wall Skyrmions that give insight into baryon density on the magnetic field and chemical potential phase diagram. At zero temperature, using the moduli effective theory, it is found Skyrmion density is saturated up to the Fermi momenta on the disk of the chiral soliton lattice orthogonal to the magnetic field. We further study chiral perturbation theory coupled to quantum electrodynamics leading to finite temperature screening, wherein a static fermionic domain-wall Skyrmion screening is predicted through a combined Debye mass/length.
}

\begin{document}

\maketitle

\section{Introduction}

The phase diagram of chromodynamics (QCD) in extreme conditions such as in a strong electromagnetic magnetic field, baryon chemical potential, and or rapid rotation remains an active venue of study due to its importance in heavy-ion collisions and neutron stars~\cite{Fukushima:2010bq,Schmidt:2017bjt,Fischer:2018sdj}. First principle calculations of QCD are fraught with technical challenges; this is particularly the case at finite density, for inclusion of a baryon chemical potential brings about the sign problem, which mar the usage of lattice QCD, one of our most widely honed computational techniques. However, low-energy dynamics are well-captured using chiral perturbation theory (ChPT)~\cite{Gell-Mann:1960mvl,Weinberg:1978kz,Gasser:1983yg,GASSER1985465,Leutwyler:1993iq}, wherein pionic degrees of freedom are employed, which demonstrate the same symmetries and their breaking within QCD. ChPT is built perturbatively about the pion dynamics to a power counting scheme in momenta~\cite{Brauner:2021sci}, but ChPT moreover possess only a few phenomenological parameters, namely the pion decay constant, $f_\pi$, and the mass, $m_\pi$. Further, strong magnetic fields are easily incorporated; as is a baryonic chemical potential and anomalous effects by way of the inclusion of a Wess-Zumino-Witten (WZW) term~\cite{Wess:1971yu,Witten:1983tx,Son:2007ny,Amano:2026jsv} that contains the Goldstone-Wilczek current~\cite{PhysRevLett.47.986,Witten:1983tx,Son:2007ny}. 

Further, in the presence of a strong magnetic field in QCD, or rather in ChPT, the ground state is predicted as a stack of domain walls orthogonal to the magnetic field in what is known as the chiral soliton lattice (CSL) of the neutral pion $\pi_0$~\cite{Son:2007ny,Eto:2012qd,Brauner:2016pko}. CSLs exist in not only two-flavor QCD, but also e.g., chiral magnets~\cite{dzyaloshinskii1964theory,Moriya:1960zz,togawa2012chiral,KISHINE20151,PhysRevB.97.184303,PhysRevB.65.064433,Ross:2020orc}
and cholesteric (chiral) liquid crystals~\cite{DEGENNES1968163}. They may also appear under thermal fluctuations~\cite{Brauner:2017uiu,Brauner:2017mui,Brauner:2021sci,Brauner:2023ort}, as well as 
$\eta'$-CSLs in 
rapid rotation~\cite{Huang:2017pqe,Nishimura:2020odq,Chen:2021aiq,Eto:2021gyy,Eto:2023tuu,Eto:2023rzd}, in addition to under a magnetic field. 
Further studies include a relation 
between CSL and Skyrmions~\cite{Kawaguchi:2018fpi,Chen:2021vou,Chen:2023jbq, Amari:2025twm},
an incommensurate crystal consisting of 
a mixed CSL of $\eta'$ and $\pi_0$ \cite{Qiu:2023guy},
the formation of CSLs 
through quantum tunneling 
\cite{Eto:2022lhu,Higaki:2022gnw} and 
dynamical formation \cite{Eto:2025ebz}, 
CSLs in 
QCD-like theories \cite{Brauner:2019aid}, 
supersymmetric QCD \cite{Nitta:2024xcu},  
and 
holographic QCD \cite{Amano:2025iwi,Amano:2026faw}, and coexistence or competition of charged pion vortices 
and CSL
with finite isospin chemical potential \cite{Gronli:2022cri,Qiu:2024zpg,Hamada:2025inf,Hamada:2026uec} 
(see also ref.~\cite{Mameda:2026kyp})
And if the magnetic field and/or baryon chemical potential go above a threshold, the CSL becomes unstable\footnote{
Another instability proposes
baryon crystals
\cite{Evans:2022hwr,Evans:2023hms}.
} to the appearance of domain-wall Skyrmions (DWSks) 
\cite{Eto:2025fkt,Eto:2023wul,Amari:2024fbo,Amari:2025twm}. DWSk are composed of both the CSL domain-wall and Skyrmions, and were first studied in (2+1)-dimensional models in refs.~\cite{Nitta:2012xq,Kobayashi:2013ju}, in (3+1)-dimensions in refs.~\cite{Nitta:2012wi,Nitta:2012rq,Gudnason:2014hsa,Gudnason:2014nba,Eto:2015uqa,Nitta:2022ahj}, in chiral magnets in refs.~\cite{PhysRevB.99.184412,Kuchkin:2020bkg,Ross:2022vsa,Amari:2023gqv,Amari:2023bmx,Amari:2024jxx,Gudnason:2024shv,Leask:2024dlo,Lee:2024lge,Gudnason:2025gqs}, and were even observed 
in refs.~\cite{Nagase:2020imn,Yang:2021nem}. 
A key finding of DWSks in a finite chemical potential and magnetic field is that for a \textit{full} period encompassing the CSL on the effective theory in the moduli approximation, the lumps supported on $\pi_2(\mathbb{C}P^1)\simeq\mathbb{Z}$ appear minimally as two baryons, and hence present with bosonics statistics~\cite{Amari:2024mip}. However, it was recently found in ref.~\cite{Copinger:2025rpo} that one may construct an effective theory over a \textit{half} period of the CSL in which the lumps appear minimally as a single baryon, and with fermionic statistics. 

The effective theory over the half period, however, does not change the associated phase diagram in comparison with our understanding of the one over the full period. Yet, since we can infer that each Skyrmion carries with it a topological charge equal to its baryon number (and indeed even an electromagnetic charge proportional to the topological charge~\cite{Amari:2024fbo}), this begs the question as to what density of Skyrmions are supported on the CSL, or for that matter what is the baryon density of (two-flavor) QCD in a magnetic field? This question poses a further complication in that since the DWSk possess an electromagnetic charge one would anticipate a competing screening of the Skyrmions. 

In this paper we address this question. Since the minimal DWSk is subject to fermionic statistics we first analyze a simple model of a fermionic quantum gas in which the fermionic Skyrmions fill a Fermi disk, due to the reduced (2+1)-dimensional CSL. Then we further incorporate a finite temperature and analyze further contributions coming from quantum electrodynamics(QED) as well as from ChPT. We also show how to incorporate a Skyrmion electromagnetic charge and subsequent screening, offering a self-consistent picture of the baryon density in low-energy QCD. From an intuitive physical standpoint, our setup of DWSks on the CSL shares some similarities to (3+1)-dimensional electromagnetic screening on the (2+1)-dimensional graphene sheet; e.g. see refs.~\cite{wunsch2006dynamical,hwang2007dielectric}.

This paper is organized as follows: In sec.~\ref{sec:chptqed} a theory combining both ChPT for two flavors and QED is outlined. We also review how the CSL emerges from a strong electromagnetic magnetic field. In sec.~\ref{sec:fermionic} we review how fermionic DWSk arise from the CSL, and also explore the various charged constituents of our combined theory. Then in sec.~\ref{sec:zero} we show how a baryonic density is predicted at zero temperature. And in sec.~\ref{sec:finite} we extend our understanding to finite temperature and calculate the Debye masses also from ChPT and QED, and show the augmented screening in such a case. Finally, in sec.~\ref{sec:conclusions} we offer conclusions.

\section{ChPT + QED effective theory with magnetic field}
\label{sec:chptqed}

ChPT~\cite{Gell-Mann:1960mvl,Weinberg:1978kz,Gasser:1983yg,GASSER1985465,Leutwyler:1993iq} serves to capture the low-energy dynamics of QCD through a theory that possesses its observed symmetries, and their breaking characteristic, with a minimal number of phenomenological parameters, namely with the pion mass and pion decay constant, $m_{\pi}$ and $f_{\pi}$ respectively. However, while electrodynamic effects are incorporated into ChPT through the presence of a WZW term~\cite{Wess:1971yu,Witten:1983tx,Son:2007ny,Amano:2026jsv}--one which we will employ--and a minimal coupling in the covariant derivatives acting on the pion fields, to fully capture the electromagnetic screening one should include effects coming from QED as well. Further, even without a finite chemical potential, a screening length can still be inherited from finite temperature. Let us first demonstrate how to simply incorporate QED into ChPT.

\subsection{ChPT + QED}

The addition of QED into ChPT is straightforward in principle. However, let us first remark on each theories range of applicability. ChPT is a low-energy description of QCD (we approximate to a two SU$(2)$ flavor pion setup, and also truncate terms in the effective description to $\mathcal{O}(p^{2})$ with exception to the WZW term\textendash see c.f., similar approaches in~\cite{Brauner:2021sci,Eto:2023wul}, but this low-energy description of quarks does not impact the electrons and their dynamics. Indeed the background electromagnetic gauge coupling often studied in ChPT under a strong magnetic field need not be weak. We will treat both a background electromagnetic field 
\begin{equation}\label{eq:mag_field}
    A_{\text{bg}\mu}(x)=\frac{B}{2}(-\delta_{\mu}^{2}x^{1}+\delta_{\mu}^{1}x^{2})\,,
\end{equation} 
where we have went ahead and specialized to a homogeneous magnetic field pointing in the $\hat{x}_{3}$ direction, and a dynamical field, $a_{\mu}$; both are coupled as 
\begin{equation}\label{eq:A}
    A_{\mu}=A_{\text{bg}\mu}+a_{\mu}\,.
\end{equation}
The dynamical field we will nevertheless treat to $\mathcal{O}(e^{2})$ in its coupling, as this will be appropriate for the evaluation of the Debye length.

Our governing Lagrangian incorporates ChPT with a WZW coupling and QED effects: $\mathcal{L}=\mathcal{L}_{\text{ChPT}}+\mathcal{L}_{\text{WZW}}+\mathcal{L}_{\text{QED}}$
with pion kinetic and mass term
\begin{equation}\label{eq:LChPT}
\mathcal{L}_{\textrm{ChPT}}=\frac{f_{\pi}^{2}}{4}\mathrm{tr}\bigl(D_{\mu}\Sigma D^{\mu}\Sigma\bigr)-\frac{f_{\pi}^{2}m_{\pi}^{2}}{4}\bigl(2-\Sigma-\Sigma^{\dagger}\bigr)\,;
\end{equation}
the pion fields are characterized through $\Sigma=\exp(i\tau^{a}\varphi^{a}/f_{\pi})=f_{\pi}^{-1}(\sigma+i\tau^{a}\pi^{a})$,
where one has the constraint $\sigma^{2}+\pi^{a}\pi^{a}=f_{\pi}^{2}$. The covariant derivative here acting on the pion fields reads $D_{\mu}\Sigma=\partial_{\mu}\Sigma+ieA_{\mu}[Q,\Sigma]$
with charge generator of the gauge group $Q=\tau_{3}/2+\mathbf{1}_{2}/6$ which dictates how the U$(1)$ electromagnetic gauge transforms; i.e., $\Sigma\to\e^{iQ\alpha(x)}\Sigma\e^{-iQ\alpha(x)}$
for a phase, $\alpha$, such that the charged pion transform as $\pi^\pm\to\e^{\pm i\alpha(x)}\pi^\pm$ with electromagnetic gauge $A_\mu\to A_\mu -\frac{1}{e}\partial_\mu \alpha(x)$.

In addition to the ChPT Lagrangian we treat a coupling to the anomalous sector of low-energy QCD by way of the WZW term. This is given as
\begin{equation}\label{eq:LWZW}
    \mathcal{L}_{\textrm{WZW}}=-(A_{\mu}^{B}+\frac{e}{2}A_{\mu})j_{\textrm{GW}}^{\mu}\,.
\end{equation}
where the Goldstone-Wilczek current~\cite{PhysRevLett.47.986,Witten:1983tx,Son:2007ny,Amano:2026jsv} is
\begin{equation}
    j_{\textrm{GW}}^{\mu}=-\frac{\epsilon^{\mu\nu\alpha\beta}}{24\pi^{2}}\mathrm{tr}\bigl(L_{\nu}L_{\alpha}L_{\beta}-3ie\partial_{\nu}[A_{\alpha}Q(L_{\beta}+R_{\beta})]\bigr)|\,,
\end{equation}
with $L_\mu=\Sigma\partial_\mu\Sigma^\dagger$ and $R_\mu=(\partial_\mu\Sigma^\dagger)\Sigma$. Notice that in addition to the U${}_B(1)$ baryon gauge field, we also have a coupling to the electromagnetic gauge field with dynamical component~\cite{Amari:2024fbo}. In this work we will consider the case of a constant baryon chemical potential
\begin{equation}
    A_\mu^B=\mu_B\delta_\mu^0\,.
\end{equation}

Finally, we will additionally consider the effects coming from QED on ChPT. We will discover there is a screening of the Skyrmions from electrons/positrons at a finite temperature. The addition of QED to the Lagrangian is
\begin{equation}\label{eq:L_QED}
    \mathcal{L}_{\text{QED}}=\bar{\psi}(i\slashed{D}-m_{e})\psi-\frac{1}{4}F_{\mu\nu}F^{\mu\nu}\,.
\end{equation}
Here the covariant derivative reads $D_\mu=\partial_\mu+ieA_\mu$ where the gauge field, eq.~\eqref{eq:A}, includes the dynamical component. $m_e$ is the electron mass. Let next review the ground state of ChPT in a magnetic field, where a chiral soliton lattice forms.

\subsection{Chiral soliton lattice in QCD}

In a strong magnetic field the two flavor QCD ChPT exhibits sine-Gordon soliton solutions. To show this transformation it is sufficient to consider the case with only the neutral pion where
\begin{equation}\label{eq:Sigma0}
    \Sigma\to\Sigma_0=\exp(i\tau_3\chi_3)
\end{equation}
becomes diagonal. Inserting the above into $\mathcal{L}_{\text{ChPT}}+\mathcal{L}_{\text{WZW}}$ using eq.~\eqref{eq:LChPT} and eq.~\eqref{eq:LWZW} we find the Lagrangian reduces to 
\begin{equation}
    \mathcal{L}\supset -\frac{f_\pi}{2}(\partial_3\chi_3)^2
    -f_\pi^2 m_\pi^2 (1-\cos\chi_3)+\frac{eB\mu_B}{4\pi^2} \partial_3\chi_3\,.
\end{equation}
The equation of motion in $\chi_3$ evaluates to $\partial_3^2\chi_3-m^2_\pi\sin\chi_3=0$ whose solution is characterized as sine-Gordon solitons, a domain-wall,
\begin{equation}\label{eq:sineG}
    \chi_{3}=2\mathrm{am}\Bigl(\frac{m_{\pi}z}{\kappa},\kappa\Bigr)+\pi\,.
\end{equation}
$\kappa$ here is the elliptic modulus, and is bound to $0\leq\kappa\leq 1$. Throughout we interchangeably use $\boldsymbol{x}=(x,y,z)$ The phase associated to the CSL is periodic about 
\begin{equation}
     \ell=\frac{2\kappa K(\kappa)}{m_\pi}\,,
\end{equation}
for $\chi_3(z)=\chi_3(z+\ell)-2\pi$.

Let us next review the effective 2-dimensional Hamiltonian integrated over one period of the CSL corresponding to above Lagrangian. We evaluate it about the sine-Gordon soliton in eq.~\eqref{eq:sineG} to find for the tension over one period as
\begin{equation}
    \sigma_\textrm{CSL} = 4m_\pi f_\pi^2 \Bigl[ \frac{2E(\kappa)}{\kappa} +\Bigl( \kappa-\frac{1}{\kappa} \Bigr) K(\kappa)\Bigr]-\frac{eB\mu_B}{2\pi}\,.
\end{equation}
From here we can determine the constraint on the elliptic modulus to be
\begin{equation}
    \frac{E(\kappa)}{\kappa}=\frac{eB\mu_{B}}{16\pi f_{\pi}^{2}m_{\pi}}\label{eq:kappa_constraint}\,,
\end{equation}
found by minimizing $\sigma_\textrm{CSL}$ over $\kappa$. The above formulae is important to have available for context of screeing of the fermionic domain-wall Skyrmions.

\section{Fermionic domain-wall Skyrmions with gauge field dynamics}
\label{sec:fermionic}

In this section let us review how fermionic DWSk arise on the CSL. Then we qualitatively analyze all the electromagnetic charged constituents present in our theory, and how they may lead to a screening of the DWSks. 

\subsection{Moduli effective theory}

DWSks are built around the standard Manton moduli approximation~\cite{MANTON198254,Eto:2006uw}. Namely one expands about the CSL, eq.~\eqref{eq:Sigma0}, to form the charged pions through $\Sigma_0\to\Sigma=g\Sigma_0g^\dagger$, where $g\in$ SU$(2)$ rotates away from the neutral pion. Then let us rewrite the pions with the two-component complex vector, $\phi$, such that
\begin{equation}\label{eq:moduli}
    \Sigma=\exp\Bigl( 2i\phi \phi^\dagger \chi_3 \Bigr)u^{-1}\,,
\end{equation}
where $\chi_3$ is the CSL sine-Gordon soliton, eq.~\eqref{eq:sineG}, and $u=\exp(i\chi_3)$. The moduli are constrained such that $\phi^\dagger \phi=1$ and $g\tau_3g^\dagger=2\phi\phi^\dagger-\boldsymbol{I}_2$, and thus the moduli live in the coset space $\mathcal{M}\cong\text{SU}(2)_V/\text{U}(1)_3\cong \mathbb{C}P^1$ owning to the ambiguity about the center charge phase. One may then re-express the pion field as
\begin{equation}\label{eq:moduli}
    \Sigma=\Bigl[ \boldsymbol{I}_2+(u^2-1)\phi\phi^\dagger \Bigr]u^{-1}\,.
\end{equation}

Next, we may construct the effective theory by introducing the moduli, eq.~\eqref{eq:moduli}, into the ChPT, eq.~\eqref{eq:LChPT}, and WZW, eq.~\eqref{eq:LWZW}, components of the Lagrangian. One may find after a few steps~\cite{Copinger:2025rpo} that ChPT half becomes
\begin{equation}
    \mathcal{L}_{\textrm{ChPT}}=\frac{f_{\pi}^{2}}{2}|1-u^{2}|^{2}\bigl[\bigl((\phi^{\dagger}D_{\alpha}\phi)^{2}+(D_{\alpha}\phi)^{\dagger}D^{\alpha}\phi\bigr)\bigr]-\mathcal{H}_\textrm{ChPT}\,,
\end{equation}
where the covariant derivative reads $D_{\mu}=\partial_{\mu}+ieA_{\mu}\tau_{3}/2$, and the corresponding effective Hamiltonian is
\begin{equation}
        \mathcal{H}_\textrm{ChPT} =\frac{f_{\pi}^{2}}{2}(\partial_{z}\chi_{3})^{2}+f_{\pi}^{2}m_{\pi}^{2}(1-\cos\chi_{3})\,.
\end{equation}
And the WZW half becomes $\mathcal{L}_\textrm{WZW}=-(A_0^B+(e/2)A_0)j^0_\textrm{GW}$, where the Goldstone-Wilczek density is now
\begin{align}
    j^0_\textrm{GW}&=-\frac{1}{2\pi}\Bigl(u-\frac{1}{u}\Bigr)^{2}\partial_{z}\chi_{3}\,q(x,y)\\
    &\quad-\frac{e\mu_{B}B}{4\pi^{2}}(\phi^{\dagger}\tau_{3}\phi)\partial_{3}\chi_{3}-\frac{e\mu_{B}}{8\pi^{2}}\epsilon^{03jk}|1-u^{2}|^{2}\partial_{3}\chi_{3}A_{j}\partial_{k}(\phi^{\dagger}\tau_{3}\phi)
\end{align}
with $q(x,y)$ being the $\mathbb{C}P^{1}$ lump topological charge density, whose integration on the 2 spatial d.o.f. of the CSL leads to $\int dxdy\,q(x,y)=k\in\mathbb{Z}$.

Now we construct the effective (2+1)-dimensional theory by integrating out $z$, the coordinate parallel to the magnetic field, and orthogonal to the CSL. To extract the \textit{fermionic} DWSk one must integrate over the half period of the CSL~\cite{Copinger:2025rpo}, i.e., over the interval $0<z<\ell/2$ or $\ell/2<z<\ell$ for e.g. the first period in the sine-Gordon soliton, eq.~\eqref{eq:sineG}. For simplicity in the steps to follow we will confine our attention to the case of $0<z<\ell/2$; analogous findings are present for the other half. How one may integrate out $z$ in the presence of a dynamical gauge field has been explored in ref.~\cite{Amari:2024fbo}, and we follow along such arguments here for the analogous case of fermionic DWSk. Namely for the gauge dynamical d.o.f. for the DWSk we assume an average over the $z$ coordinate such that $a(x^\mu)$ where now $\mu = (0,1,2)$. Quadratic fluctuations in $a$ in eq.~\eqref{eq:A} can indeed influence the screening characteristics, and we will revisit the issue for the finite temperature case--where we keep all dynamical d.o.f. present to $\mathcal{O}(p^2)$. Since we take our effective DWSk theory with $a(x^\mu)$ with $\mu = (0,1,2)$, we may follow along the steps in ref.~\cite{Copinger:2025rpo} to find the effective theory.

Let us first integrate out $z$ for the ChPT Lagrangian. We find the effective theory may be expressed in terms of the covariant derivative weighted over the K\"ahler class:
\begin{equation}
    \int^{\ell/2}_0dz\,\mathcal{L}_\textrm{ChPT}=\mathcal{C}(\kappa)\frac{1}{2}\bigl[(\phi^{\dagger}D_{\alpha}\phi)^{2}+(D_{\alpha}\phi)^{\dagger}D^{\alpha}\phi\bigr]-\sigma_\textrm{ChPT}\,,
\end{equation}
where one has that 
\begin{equation}
    \mathcal{C}(\kappa) =\frac{16f_{\pi}^{2}}{3m_{\pi}}\frac{1}{\kappa^{3}}\Bigl[(2-\kappa^{2})E(\kappa)+2(\kappa^{2}-1)K(\kappa)\Bigr]\,.
\end{equation}
And we also have that the tension becomes over the half effective theory
\begin{equation}
    \sigma_\textrm{ChPT}\coloneqq \int^{\ell/2}_0dz\,\mathcal{H}_\textrm{ChPT}=2f_{\pi}^{2}m_{\pi}\Bigl[\frac{2}{\kappa}E(\kappa)+\Bigl(\kappa-\frac{1}{\kappa}\Bigr)K(\kappa)\Bigr]\,.
\end{equation}
Let us next treat the effective domain-wall contributions coming from the WZW term; for fermionic half effective theory,~\cite{Copinger:2025rpo}, we find that
\begin{equation}
    \int^{\ell/2}_0dz\,\mathcal{L}_\textrm{WZW}=-\mu_B q-\frac{e\mu_{B}}{4\pi}\epsilon^{03jk}\partial_{j}(A_{k}\phi^{\dagger}\tau_{3}\phi)\,.
\end{equation}
The first term, the Skyrmion topological charge density, stems from 
\begin{equation}
    \mathcal{B}=\frac{-1}{24\pi^{2}}\epsilon^{ijk}\mathrm{tr}[L_{i}L_{j}L_{k}]
    =-\frac{1}{2\pi}\Bigl(u-\frac{1}{u}\Bigr)^{2}\partial_{z}\chi_{3}q(x,y)\,.\label{eq:beta}
\end{equation}
in the WZW Lagrangian under the moduli approximation such that $\int^{\ell /2}_0 dz \mathcal{B}=q$. We have an effective DWSk theory, however, it is convenient to rather express the  $\mathbb{C}P^{1}$ theory as one of a $O(3)$ nonlinear sigma model.

Since one has that $\mathbb{C}P^{1}\simeq S^2\simeq O(3)/O(2)$, we may re-expression the pion phase field, $\phi$, in terms of the three coordinates
\begin{equation}
    \vec{n}=\phi^\dagger \vec{\tau}\phi=\mathrm{tr}(\vec{\tau}P)\,,
\end{equation}
where $P=\phi \phi^\dagger = (1/2)(\boldsymbol{I}_2+\vec{n}\cdot \vec{\tau})$. One has that $\vec{n}^2 =1$. The neutral pion follows from $n_3$, and the charged pions $\pi^\pm\propto n_1 \mp in_2$. Using the above map one may express $\mathcal{L}_\textrm{DW}=\int^{\ell/2}_0dz\,\mathcal{L}_\textrm{ChPT}+\int^{\ell/2}_0dz\,\mathcal{L}_\textrm{WZW}$ as an $O(3)$ nonlinear sigma model
\cite{Amari:2024fbo}:
\begin{equation}\label{eq:DW_Lagrangian}
\mathcal{L}_{\textrm{DW}} =\frac{\mathcal{C}(\kappa)}{8}D_{\alpha}\vec{n}\cdot D^{\alpha}\vec{n}-\sigma_\textrm{ChPT}-(\mu_{B}+\frac{e}{2}A_{0})\Bigl\{ q-\frac{e}{4\pi}\epsilon^{03jk}\partial_{j}[A_{k}(1-n_{3})]\Bigr\}\,,
\end{equation}
where the lump charge can be re-expressed in terms of $\vec{n}$ as $q=(8\pi)^{-1}\epsilon^{ij}\vec{n}\cdot(\partial_{i}\vec{n}\times\partial_{j}\vec{n})$.
And the covariant derivatives now read $D_\mu (n_1+in_2)=(\partial_\mu-ieA_\mu)(n_1+in_2)$ for the charged pions, and $D_\mu n_3=\partial_\mu n_3$ for the neutral pions. We remark that the covariant derivative acting on the $O(3)$ variables differs from the definition employed for QED where one has that $\partial_\mu+ieA_\mu$.

\subsection{Charged constituents in DWSk + QED}

Having constructed the DWSk effective theory let us explore its electromagnetically charged constituents; we will also analyze those coming from QED. We begin with the former. Since the domain-wall effective theory is in (2+1)-dimensions, we can find the charge surface area on the CSL as
\begin{align}\label{eq:charge_carriers}
\sigma_{\textrm{DW}} & =\delta/\delta a^{0}\int dx^{0}d^{2}x\,\mathcal{L}_{\textrm{DW}}\notag\\
 & =\frac{e\mathcal{C}(\kappa)}{4}\boldsymbol{n}\cdot iD_{0}\boldsymbol{n}-\frac{e}{2}q+\frac{e^{2}}{8\pi}\epsilon^{03jk}\partial_{j}[A_{k}(1-n_{3})]\,.
\end{align}
Importantly we find that the topological charge density indeed also possesses an electro\-magnetic charge density, namely~\cite{Amari:2024fbo} for lump $k$ where $\int d^2x q(x,y)=k$ one has the charge coupling $-e/2$ for each unit of $k$ topological charge--that we associate to the baryon number. The screening of these $k$ lumps is the principal target of our analysis as we can determine equilibrium baryon densities from it.

One also has that the charged pions can induce a charge current associated with the kinetic term in eq.~\eqref{eq:charge_carriers}. However, there are two caveats here: 1. Notice that the kinetic term in the effective Lagrangian, eq.~\eqref{eq:DW_Lagrangian} is quadratic in the $O(3)$ nonlinear sigma coordinates--indeed this is the case for the full theory before integrating out $z$ in eq.~\eqref{eq:LChPT}. What this tells us is that upon integrating out the fields for the kinetic contributions, $\vec{n}$, or $\Sigma$ for that matter, and producing the functional determinant, due to Wick's theorem, odd powers of $a^\mu$ vanish including the linear contribution. Thus we expect for the kinetic contributions, leading order corrections in $a^\mu$ will enter at the quadratic order, indicative of a Debye screening. 2. As just mentioned, the full theory predicts fluctuations in $a^\mu$ along the $z$ direction as well--this is only approximately captured in the effective description given in eq.~\eqref{eq:DW_Lagrangian}. We will treat both points 1 and 2 with rigor in a section to follow by calculating the effective action stemming from the kinetic term of ChPT, and show its quadratic $\mathcal{O}(a_0^2)$ form.

The CSL term, $(e^{2}/8\pi)\epsilon^{03jk}\partial_{j}[A_{k}(1-n_{3})]$, does indeed also possess a charge, and so does the CSL as explored in ref.~\cite{Son:2007ny}. This is reflected by the background field part of the CSL term, $(e^{2}/8\pi)\epsilon^{03jk}\partial_{j}[A_{\text{bg}k}(1-n_{3})]$ where $A_\text{bg}$ is the background magnetic field, eq.~\eqref{eq:mag_field}. It provides a uniform charge spread over the CSL, but will not influence a screening of the DWSk. Consider the effective theory in eq.~\eqref{eq:DW_Lagrangian} coupled to the Maxwell Lagrangian in eq.~\eqref{eq:L_QED}. Upon evaluating the inhomogeneous Lorentz equation quadratic contributions from e.g. the ChPT part will induce a screening but linear parts as in the background CSL are only source terms and cannot induce an effective photon mass term. (One does have the neutral pion, $n_3$, dependence in the CSL term, however, term is excluded in minimal energy lump configurations; see refs.~\cite{Eto:2023wul,Eto:2025fkt}.) Also one has a quadratic term in $a^\mu$: $(e^{2}/8\pi)a_0\epsilon^{03jk}\partial_{j}[a_{k}(1-n_{3})]$ as it sits in the domain-wall Lagrangian in eq.~\eqref{eq:DW_Lagrangian}. This predicts a perpendicular Debye length being proportional to $a_0a_k$. Since we are primarily concerned with the static screening of DWSks coming from $a_0^2$ terms we will not further study such a contribution.

Finally, we have a (3+1)-dimensional current coming from the QED sector, namely $\bar{\psi}\gamma_\mu\psi$. Then along the same lines as argued above for the ChPT kinetic term, one will only see a Debye length here associated to the $\mathcal{O}(a_0^2)$ contributions in the QED effective action. 

\begin{figure}
\centering
\includegraphics[scale=0.3]{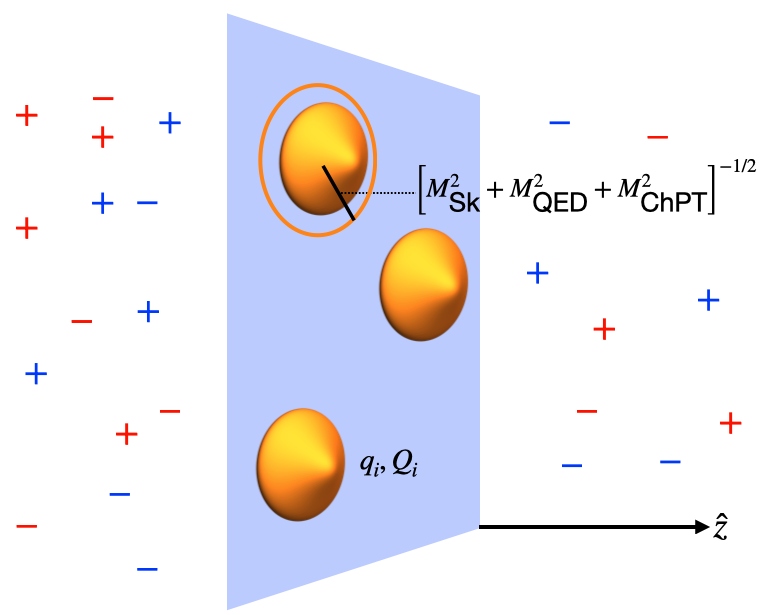}
\caption{Charged constituents in the DWSk + QED theory. DWSks are the yellow half macaron-like lumps, which posses a topological charge, $q_i$, and electromagnetic charge, $Q_i$ for specie $i$. They are confined to the CSL, the blue surface. In the bulk a Debye screening is present coming from both the ChPT and QED sectors, represented figuratively with `+' and `-' charge distributions, respectively in blue and red font. The combined screening length for a DWSk is given by $[M_\textrm{Sk}^2+M_\textrm{QED}^2+M_\textrm{ChPT}^2]^{-1/2}$, and is figuratively represented with the orange boundary about the DWSk isosurface. }
\label{fig:charges}
\end{figure}

All in all one would expect a Debye screening coming from the Skyrmions themselves, the kinetic ChPT, and QED sectors; we may associate the screening to Debye masses $M_\textrm{Sk}$, $M_\textrm{ChPT}$, and $M_\textrm{QED}$. And we would expect charged lumps, the fermionic DWSks, sitting on a charged CSL. The physical scenario is depicted in fig.~\ref{fig:charges}.
There half macaron-like fermionic DWSks (yellow) sit on the CSL (blue surface) and are screened by themselves, and from contributions from QED and ChPT, represented figuratively with `+' and `-' charges in red and blue respectively. The combined screening length is given by the inverse length $[M_\textrm{Sk}^2+M_\textrm{QED}^2+M_\textrm{ChPT}^2]^{-1/2}$, that we will now go onto determine after showing how baryonic Skyrmions appear on the domain-wall.

We will go onto calculated the DWSk density at both zero temperature and at finite temperature, and as is known for the Skyrmions themselves that negatively charged Skyr\-mions associated with anti-lumps are energetically favorable. According to eq.~\eqref{eq:charge_carriers} and with the QED charge density in mind, there must be a compensating positive electromagnet\-ically charged density to pair against the negatively charged Skyrmions that respects a charge neutrality of the closed system. We go onto calculate the DWSk density, which is screened and which is of interest here; however, a compensating charge density would be present elsewhere in our setup of ChPT with QED, and we assume its presence, but we only treat the DWSk density directly.

\subsection{Fermionic domain-wall Skyrmions}

Before constructing a domain-wall effective energy and the minimal energy profile for the DWSk, let us comment and review on the DWSk's fermionic nature. This is an essential point in this work, as the DWSk fermionic statistics govern their screening nature and equation of state. The identification of the minimal DWSk as fermions with baryon number one was examined in ref.~\cite{Copinger:2025rpo}. There most notably in contrast to the full effective theory built from an integration over the full CSL period from e.g. $0\leq z\leq \ell$, one may break up the period into halves ($0\leq z\leq \ell/2$ and $\ell/2\leq z\leq \ell$) each identifiable with single baryon charge for the minimal lump profile. This was largely identified with the topological charge as shown with $\int d^2x q(x,y)=k$. It was also further shown using the Witten method~\cite{Witten:1983tx}, that the DWSks indeed obeyed fermionic statistics by using an embedding of SU$(3)$ onto the SU$(2)$ pion field to study a five dimensional WZW term that predicted spin statistics; see ref.~\cite{Copinger:2025rpo,Amari:2024mip} for further details. Last it was determine that the phase diagram for the fermionic DWSks was identical to that of the bosonic one over the full period, which we now go onto show.

In ref.~\cite{Amari:2024fbo} the bosonic doubled theory of eq.~\eqref{eq:DW_Lagrangian} was explored with gauge field dynamics; it was found that the Skyrmions found from the Bogomol'nyi-Prasad Sommerfield (BPS) approximation (in which gauge field dynamics are neglected) possess exactly the same phase boundary structure as Skyrmions that fully incorporate gauge field dynamics. Since the half fermionic theory and full bosonic theory also possess the same phase structure, we can conclude that gauge field dynamics will not affect the appearance of fermionic DWSks, and the gauge field dynamics can be neglected in the discussion to follow in this section. Thus the avenue to the fermionic DWSk is the same as demonstrated in ref.~\cite{Copinger:2025rpo}, and let us review pertinent details in this subsection. Even so, we must still take that the Skyrmions have electromagnetic charge proportional to their topological charge, namely $-e/2$ for each unit of $k$ topological charge.

To begin let us examine the static energy profile of the domain-wall Lagrangian by determining the corresponding Hamiltonian in the absence of the dynamical gauge; this is up to a constant and hence omitting $\sigma_\textrm{ChPT}$
\begin{equation}\label{eq:HDW}
    \mathcal{H}_{\textrm{DW}} =\mathcal{C}(\kappa)\frac{1}{8}[\partial_{i}\boldsymbol{n}\cdot\partial_{i}\boldsymbol{n}]+\mu_{B}q-\frac{e\mu_{B}}{4\pi}\epsilon^{03jk}\partial_{j}[A_{k}(1-n_{3})]\,.
\end{equation}
Then we look to find the accompanying energy bound. This can be accomplished with the BPS approximation. The (anti-)BPS equation is $\partial_i \boldsymbol{n}\pm\epsilon_{ij}\boldsymbol{n}\times \partial_j \boldsymbol{n}=0$. Solutions to the BPS equation will furnish (anti-)BPS lumps~\cite{Eto:2023wul,Eto:2025fkt}, splitting energies into either lumps with $k>0$ or anti-lumps with $k<0$ with
\begin{equation}
    k=\int d^2x \,q(x,y)\in\pi_2(\mathbb{C}P^1)\,.
\end{equation}
Solutions to the BPS equation may be characterized by a set of complex parameters $\{a_A,b_A\}$ with $A=0,1,...,|k|-1$ that serve as moduli parameters; the form of the BPS solutions is not important for our analyses to come, but concrete expressions may be found in~\cite{Eto:2023wul,Eto:2025fkt}. Applying the BPS approximation to~\eqref{eq:HDW} one may find for the domain-wall effective energy for a lump the following~\cite{Copinger:2025rpo}\:
\begin{equation}\label{eq:EDW1}
    E_{\textrm{DW}}\leq\pi\mathcal{C}(\kappa)|k|+\mu_{B}k+\frac{e}{2}\mu_{B}B|b_{k-1}|^{2}\,,
\end{equation}
where we have used that $\partial_i\boldsymbol{n}\cdot\partial_i\boldsymbol{n}=(1/2)[\partial_i\boldsymbol{n}\pm\epsilon_{ij}\boldsymbol{n}\times\partial_j\boldsymbol{n}]^2\pm 8\pi q$ and $\int d^{2}x\epsilon^{03jk}\partial_{j}[A_{k}(1-n_{3})]=-2\pi B|b_{k-1}|^{2}$.

Let us now examine the minimal energy profile from the above. First, the last term in eq.~\eqref{eq:EDW1} is positive semi-definite, thus by energy minimization one may take that $b_{k-1}=0$. Further, we can also see that anti-lump configuration with $k<0$ lower the energy. We have now for the minimal energy profile
\begin{equation}\label{eq:EDWmin}
E_{\textrm{DW}}^\textrm{min}=\pi\mathcal{C}(\kappa)|k|+\mu_{B}k\,.
\end{equation}
One may also determine the phase boundary by which DWSk appear; this occurs when $\mu_c=\pi\mathcal{C}(\kappa)$. The DWSk phase boundary can be seen in fig.~\ref{fig:phase} as the green line. Also, for completeness, in the phase diagram one can see the phase boundary for the the CSL to appear as the solid blue line, below which is the QCD vacuum. The dotted blue line represents the metastability of the CSL where the charged pion condensate (CPC) appears. And the red area represents the extent to which ChPT holds. For further details we refer to ref.~\cite{Copinger:2025rpo}. Surprising with only the minimal energy profile given in eq.~\eqref{eq:EDWmin} one may infer quantitative information about the zero temperature screening of DWSk, which we now go onto explore.

\section{Domain-wall Skyrmion zero temperature screening}
\label{sec:zero}

Here we treat the DWSks as a dilute relativistic Fermi gas; this informs us of the filling density on the Fermi disk and gives insight into zero temperature screening. We will later explore the additional effects of including the Skyrmion electromagnetic charge--alongside finite temperature screening from QED and ChPT. The analysis here is twofold important: First, the quantum free gas model we employ here provides a leading order approximation to Skyrmion density with physically opaque expressions. Second, in order to determine the effect of a Skyrmion electromagnetic charge we must construct a self-similar solution to the equation of state; this begins with the analysis here. 

Notice that for anti-lump configurations our energy profile is of the form of a grand canonical Hamiltonian, $H-\mu_{B}N$, where the number operator, $N=-\int d^2x q$ (or rather $n=-q$), is played by the role of the WZW topological lump charge for the anti-BPS solutions with $k<0$. Further the condensation threshold is fixed by the phase boundary at $\mu_B>\mu_c$ for the single lump, namely with $k=-1$. Further, since we have determined the \textit{static} minimal energy profile in eq.~\eqref{eq:EDWmin}, this signals that we consider a Skyrmion mass as one of the K\"ahler class with 
\begin{equation}\label{eq:mSk}
    m_{\textrm{Sk}}\coloneqq\pi\mathcal{C}(\kappa)=\mu_c\,.
\end{equation}
Now we can understand the criteria for the formation of Skyrmions as $\mu_{B}>m_{\textrm{Sk}}$ where it is energetically favorable to fill the Skyrmion Fermi sea. In this picture we see the DWSks are treated as non-interacting fermions\footnote{Treating the DWSk as a non-interacting fermionic particle is consistent with the observations that, in the BPS approximation, the DWSk experiences no static interactions and that energy minimization requires the size modulus of a single anti-lump $(k=-1)$ to vanish, namely, $b_{-2}=0$ in eq.~(\ref{eq:EDW1}).}, and thus the fermionic DWSks form a fermi gas.

\begin{figure}
\centering
\includegraphics[scale=0.9]{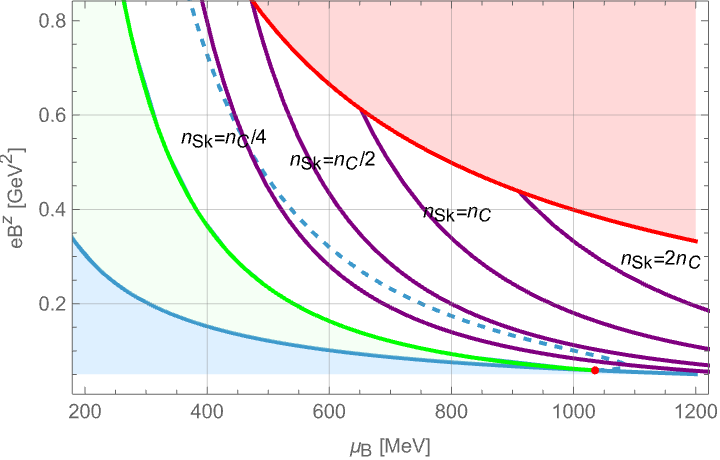}
\caption{Baryon density of the free quantum Fermi gas model, eq.~\eqref{eq:n_free} overlaid on top of the QCD phase diagram for finite chemical potential with a magnetic field. Densities are expressed in units of a nominal density $n_\textrm{C}=(\pi/m_{\pi})[8f_{\pi}^{2}/3]^{2}$, and are shown as isocurves in purple font. The green isocurve in which $n_\textrm{Sk}=0$ is also the phase boundary for the DWSk phase. The solid blue line represent the phase boundary between the QCD vacuum and the CSL, the CSL phase is the area in green and the QCD vacuum phase is the area in blue. And above the dotted blue line the phase boundary represents where the charged pion condensate (CPC) phase appears at $B\geq B_\textrm{CPC}\sim 16 \pi^4 f_\pi^4/\mu_B^2$, and in which the CSL becomes unstable~\cite{Brauner:2016pko}. The tricritical point in which all three phase boundaries meet is at $(16\pi f_\pi^2/(3m_\pi),3m_\pi^2$). The red area depicts a region in which ChPT no longer applies; see ref.~\cite{Copinger:2025rpo}. }
\label{fig:phase}
\end{figure}

Since we have treated a static configuration, we may assume a Lorentz boost of the static energy, and hence a dispersion profile follows as
\begin{equation}\label{eq:dispersion}
\varepsilon_{\textrm{Sk}}=\sqrt{m_{\textrm{Sk}}^{2}+\vec{p}^{2}}
\end{equation}
for the 2-dimensional momenta $\vec{p}$. Now a key distinction of the identification of the minimal profile of the DWSk over a half-period of the CSL obeying fermionic statistics is that we may assume an occupation number from a filling of the Fermi surface. First, we determine the Fermi momentum stemming from the dispersion relation in eq.~\eqref{eq:dispersion} as
\begin{equation}\label{eq:kF}
    k_F=\sqrt{\mu_B^2-m_\textrm{Sk}^2}\,.
\end{equation}
Then at \textit{zero temperature} the occupation number is a step function up to the Fermi momentum lying on the Fermi disk with $|\vec{p}|\leq k_F$ with all available quantum states filled. And hence we can determine a \textit{two dimensional} density of created Skyrmions as 
\begin{equation}\label{eq:nSk_def}
\sigma_{\textrm{Sk}}=\int_{|\vec{p}|\leq k_F}\frac{d^{2}p}{(2\pi)^{2}}=\frac{1}{2\pi}\int^{k_F}_0 dp\,p=\frac{k_{F}^{2}}{4\pi}\,.
\end{equation}
To determine the three dimensional density one need only introduce the half periodic length orthogonal to the CSL as $n_\textrm{Sk}=(2/\ell)\sigma_\textrm{Sk}$. Then using eq.~\eqref{eq:kF} and eq.~\eqref{eq:mSk} we can determine for the three dimensional Skyrmion density as
\begin{equation}\label{eq:n_free}
n_{\textrm{Sk}}=\frac{\mu_{B}^{2}-(\pi\mathcal{C}(\kappa))^{2}}{2\pi\ell}\,.
\end{equation}
Note that we require $\mu_{B}>\pi\mathcal{C}(\kappa)$, or else there can be no Skyrmions and the density is zero. A plot of density isocurves can be seen in fig.~\ref{fig:phase} as a function of the nominal density $n_\textrm{C}=(\pi/m_{\pi})[8f_{\pi}^{2}/3]^{2}$ on the $B$ and $\mu_B$ QCD phase diagram. To remind, the green line represents the threshold for $\mu_B>\mu_c$, and the criteria for DWSk to appear. All the density isocurves appear to the upper-right of the DWSk threshold as anticipated. With increasing baryon chemical potential the baryon density increases, also as anticipated. However, as the magnetic field strength increases we can see that the DWSk density, $n_\textrm{Sk}$ also increases; we can see this for a fixed $\mu_B$ in the figure and as we go up the density increases.

It was determined from eq.~\eqref{eq:charge_carriers} that the fermionic DWSk do in fact possess an electromagnetic charge, and thus we expect a screening of the DWSks. We can incorporate the screening nature by self-consistently solving for the density of created Skyrmions on the disk in eq.~\eqref{eq:nSk_def} using an electromagnetically density modified Fermi-momentum. To compute this we will need the corresponding Debye mass coming from the above model. This is easily found from the Skyrmion density
\begin{equation}
M_{\textrm{Sk}}^{2}=\frac{e^{2}}{4}\frac{\partial n_{\textrm{Sk}}}{\partial\mu_{B}}=\frac{e^{2}\mu_{B}}{4\pi\ell}\,.
\end{equation}
However, before outlining how to treat the Debye mass self-consistently, it is convenient at this point to find the thermal Debye masses coming from the QED and ChPT sectors of our theory, as all the Debye masses may be simply superposed together. 

Let us also remark at this point that the zero-temperature Debye mass gives a quantit\-ative estimate in which the fermi gas model applies. The Debye length is $\lambda_\textrm{Sk}=M_\textrm{Sk}^{-1}$ from which we can find a parameter cutoff, $n_\lambda\sim\lambda_\textrm{Sk}^{-3}$, beyond which screening dominates the mean field fermi gas model employed here. However, once evaluating $n_\lambda\approx n_\textrm{C}$ over a range of applicable CSL solutions as function of $\kappa$, one can discover that the isocurve representing the cutoff lies in red region of fig.~\ref{fig:phase}, and thus our model is applicable everywhere ChPT is valid. For reference the red line in fig.~\ref{fig:phase} is found from $B_\text{max}=16\pi^3 f_\pi^3/\mu_B$; see ref.~\cite{Copinger:2025rpo}. 

While the DWSk effective Hamiltonian may be interpreted as a grand canonical one of a free fermi gas without interactions, by incorporating a Debye mass and finite temperature effects to follow, interactions are no longer negligible and there is a smooth crossover to a fermi liquid. As the density increases, the mean inter-Skyrmion distance eventually becomes comparable to the Debye length. At this point the screening clouds overlap and the non-interacting Fermi-gas approximation breaks down. The DWSks are then expected to form an interacting Fermi liquid. Without the screening the above model can be viewed as a dilute limit of the DWSk system. Consider the Debye length, $\lambda_{\textrm{Sk}}$, as a scale with which the crossover to the fermi liquid becomes relevant. For $d\sim n^{-1/3}\gg \lambda_{\textrm{Sk}}$ interactions are negligible between the DWSks and a fermi gas is formed. However as $d\sim \lambda_{\textrm{Sk}}$ the interactions are no longer negligible and the Skyrmions form a fermi liquid. At even higher densities, where the Coulomb repulsion dominates over the kinetic energy, crystallization of DWSks may also become favorable.

\section{Linear response and finite temperature screening}
\label{sec:finite}

Above we discovered that even at zero temperature by virtue of the fermionic statistics of the DWSks we have an electromagnetic screening. This is also possible because the WZW term enters at finite chemical potential; indeed in ChPT and naturally in QED the photon does not acquire a mass term at $a_0^2$ in the Lagrangian when vacuum polarization effects are considered. This, however, is not the case at finite temperature where indeed at thermally generated photon mass-like term arises. We explore such corrections here in this section. 

One may consider finite temperature effects the standard way by replacing the sum over energies $p_0$ as a Matsubara sum. This step is however, nuanced in the presence of a background field, such is as required for our setup with a strong electromagnetic magnetic field and baryonic electromagnetic coupling, $A^B$. The problem is a well-studied one, and we follow the approach as outlined in refs.~\cite{Gies:1998vt,Dittrich:2000zu}, wherein gauge transformations of the background fields are taken in such a way as to preserve the (anti-)periodicity requirements of the imaginary time formalism when extending to finite temperature. Before elaborating on this point, let us first illustrate using QED as an example, how the effective action contains information on the electromagnetic screening characteristics, and how it provides a convenient way to find such information.

\subsection{QED Debye mass}
Turning our attention the QED sector let us write for our effective
action of the Lagrangian in eq.~\eqref{eq:L_QED}
\begin{equation}\label{eq:qedeff}
S_{\text{QED}}^{\text{eff}}=-\frac{i}{2}\mathrm{Tr}\ln[\slashed{D}^{2}+m_{e}^{2}]-\frac{1}{4}\int d^{4}x\,F_{\mu\nu}F^{\mu\nu}\,.
\end{equation}
Though we do not integrate out the gauge field. Rather, from an expansion in the dynamical gauge coupling, $e$, we can identify the leading order quadratic contribution at $\mathcal{O}(a^{2})$ as the definition of the Debye mass; this is formally at the Lagrangian level represented by the polarization tensor $-\frac{1}{2}\int d^{4}xd^{4}y\,a^{\mu}(x)\Pi_{\mu\nu}(x,y\,|A_{\text{bg}})a^{\nu}(y)$. At zero temperature the $\Pi_{00}$ would vanish. However, even with the extension to finite temperature one would have to embark on a laborious calculation of the polarization tensor. And we can circumvent this difficulty by instead treating a coupling to a constant finite temperature gauge $\mu$ that is analogous to a finite chemical potential ($\mu$ differs from $\mu_B$ and will instead be treated as a dummy variable). The constant $\mu$ is the first order in a derivative expansion of the one-loop effective action, eq.~\eqref{eq:qedeff}, written for finite temperature, that moreover furnishes one with the exact leading order expression of the QED Debye mass~\cite{Gies:1998vt,Dittrich:2000zu}. Let us now explain how to extend eq.~\eqref{eq:qedeff} to finite temperature and what $\mu$ means.

We follow along the derivation as presented in refs.~\cite{Gies:1998vt,Dittrich:2000zu}, and we refer the reader to the discussions therein for details of steps to follow. Refs.~\cite{Gies:1998vt,Dittrich:2000zu} have reported a Debye mass at weak fields. We show how to arrive at the extension to a strong magnetic field as is relevant for our setup. However, this is not entirely new since refs.~\cite{Gies:1998vt,Dittrich:2000zu} also have calculated the one-loop thermal corrections to the effective action, from which the Debye mass may be simply inferred. We will, however, treat with some care the one-loop correction to the ChPT effective action to follow, which is indeed a new finding--in addition to the the screening characteristics already presented in sec.~\ref{sec:fermionic}.

Consider a heat bath at $\beta=1/T$, characterized with 4-velocity vector, $u^\mu$, such that $u^2=1$ and is time-like. In going to imaginary time one must impose the anti-periodic requirements over interval $x^\mu$ to $x^\mu+i\beta u^\mu$ for the fermion d.o.f., and we may accomplish this for the QED effective action. This can be built from the zero temperature case conveniently written in Schwinger propertime~\cite{Schwinger:1951nm}:
\begin{equation} \label{eq:L1QED}
    \mathcal{L}^1_\textrm{QED}\coloneqq -\frac{i}{2}\mathrm{Tr}\ln[\slashed{D}^{2}+m_{e}^{2}]=
    \lim_{x'\to x}\frac{i}{2}\mathrm{tr}\int^\infty_0\frac{ds}{s}\langle x|e^{-i[m^2_e+\slashed{D}^2]s} |x'\rangle\,.
\end{equation}
In contrast to conventional studies of the one-loop effective action to make the connection to finite temperature one imposes a coincident limit onto the propertime kernel. For later reference let us go ahead and record the exact one-loop effective action in a magnetic field~\cite{Schwinger:1951nm}
\begin{equation}
    \mathcal{L}_{\text{QED}}^{1T}=\frac{eB}{4\pi^{2}}\int_{0}^{\infty}\frac{ds}{s^{2}}e^{-ism_{e}^{2}}\cot(eBs)\,.
\end{equation}
A simple way to derive the above is to convert the propertime kernel into a path integral, whose form is Gaussian and hence and exactly solvable; see e.g. refs.~\cite{Dunne:2004nc,UsRep,103}. Now, to arrive at the finite temperature extension one may apply the method of image sources to the Green function~\cite{Gies:1998vt,Dittrich:2000zu}; what this amounts to is an augmentation of the coincident limit so that $x'\to x'_n$ in right hand side of eq.~\eqref{eq:L1QED} where
\begin{equation}\label{eq:xshift}
    x^{\prime\mu}_n=x^{\prime\mu}-i\beta n u^\mu\,,
\end{equation}
and where $n$ denotes the Matsubara sum over $n\in\mathbb{Z}$. Then one may define for the one-loop thermal effective action
\begin{equation}\label{eq:L1TQED_def}
    \mathcal{L}_{\text{QED}}^{1T}=
    \lim_{x'\to x}\frac{i}{2}\mathrm{tr}\sum_{n=-\infty}^\infty(-1)^n\int^\infty_0\frac{ds}{s}\langle x|e^{-i[m^2_e+\slashed{D}^2]s} |x'_n\rangle\,.
\end{equation}
We acquire a $(-1)^n$ factor from the anti-periodic boundary conditions of the fermions and hence their determinant.

To evaluate eq.~\eqref{eq:L1TQED_def} and perform the Matsubara sum one may write the multiple kernel in momentum space. A key point is that the non-trivial extension of the coordinate by $-i\beta n u^\mu$ in eq.~\eqref{eq:xshift} amounts to an addition to the gauge holonomy factor present in the kernel; the addition appears as a gauge transformation. This takes the form in the kernel as the following factor: $\beta^{-1}\int^\beta_0d\tau A_\mu(x+i\tau u)u^\mu\to\mu$, and hence giving rise to a chemical potential. We refer to refs.~\cite{Gies:1998vt,Dittrich:2000zu} for details. The evaluation of the one-loop thermal effective action, eq.~\eqref{eq:L1TQED_def}, in a constant magnetic field background has been carried out in ref.~\cite{Dittrich:2000zu}; it reads
\begin{equation}
\mathcal{L}_{\text{QED}}^{1T}=\frac{eB}{4\pi^{2}}\int_{0}^{\infty}\frac{ds}{s^{2}}e^{-ism_{e}^{2}}\cot(eBs)\sum_{n=1}^{\infty}(-1)^{n}e^{i\frac{n^{2}}{4T^{2}}\frac{1}{s}}\cosh\Bigl(\frac{e\mu n}{T}\Bigr)\,.
\end{equation}
Importantly we can see that the thermal effects amount to the insertion of a propertime factor to the zero-temperature case, eq.~\eqref{eq:L1QED}. Then as reasoned above, one may perform an expansion in $a_0=\mu$ to determine the quadratic contribution to the QED effective action, which is identifiable as the quadratic Debye mass, i.e.,
\begin{equation}
\label{eq:m_Dformula}
M_\textrm{QED}^2=\frac{\partial^{2}\mathcal{L}_{\text{QED}}^{1T}}{\partial\mu^{2}}\Big|_{\mu\to0}\,.
\end{equation}
Such an identification is valid for constant $a_0=\mu$ because one may identify it as the leading (zeroth) order expansion of a derivative expansion of the inhomogeneous and dynamical $a_0(x)$. A weak-field expansion of the inhomogeneous $a_0(x)=\mu(x)$ one-loop thermal effective action takes the form $\mathcal{L}^{1T}_\textrm{QED}=-(1/2)\partial_\mu \mu(x)\partial^\mu \mu(x)+(M_\textrm{QED}^2/2)\mu^2(x)+\mathcal{O}(\mu^4(x))$~\cite{Gies:1998vt,Dittrich:2000zu}.

From the one-loop thermal effective action above, we can determine the Debye mass directly  using eq.~\eqref{eq:m_Dformula} to find
\begin{align}
M_\textrm{QED}^2 & =\frac{eB}{4\pi^{2}}\int_{0}^{\infty}\frac{ds}{s^{2}}e^{-ism_{e}^{2}}\cot(eBs)\sum_{n=1}^{\infty}(-1)^{n}e^{i\frac{n^{2}}{4T^{2}}\frac{1}{s}}\Bigl(\frac{en}{T}\Bigr)^{2}\\
 & =-\frac{eB}{\pi^{2}}\frac{e^{2}}{T}\sum_{n=1}^{\infty}(-1)^{n}n\Biggl\{ m_{e}K_{1}\Bigl(\frac{nm_{e}}{T}\Bigr)+2\sum_{l=1}^{\infty}\sqrt{m_{e}^{2}+2leB}K_{1}\Bigl(\frac{n\sqrt{m_{e}^{2}+2leB}}{T}\Bigr)\Biggr\}\,,
\end{align}
where we have expanded about the Landau levels in the cotangent term with the magnetic field in the first line, and then completed the propertime integrals to arrive at a sum over Bessel functions over Landau levels and Matsubara number $n$. The above is valid for arbitrary temperature and magnetic field. However,  since temperature modifies the standard power counting scheme present in ChPT as $\partial_\mu,m_\pi,T,A_\mu=\mathcal{O}(p^1)$ for characteristic momentum $p$~\cite{Brauner:2021sci}, we cannot reasonably assume Debye masses much larger than the pion mass hold in our model. Therefore, we treat the low temperature case of the above. We can also readily evaluate the expression in the low temperature limit since the modified Bessel functions exponentially damp in such a regime. We find to leading order
\begin{equation}
M_\textrm{QED}^2  \sim-\frac{eB}{\pi^{2}}\frac{e^{2}}{\sqrt{T}}\sum_{n=1}^{\infty}(-1)^{n}n\Biggl\{\sqrt{\frac{\pi m_{e}}{2n}}e^{-\frac{nm_{e}}{T}}+2\sum_{l=1}^{\infty}\sqrt{\frac{\pi\sqrt{m_{e}^{2}+2leB}}{2n}}e^{-\frac{n\sqrt{m_{e}^{2}+2leB}}{T}}\Biggr\}\,,
\end{equation}
which will clearly vanish in the $T\to0$ limit. Then for $T\ll \sqrt{m_e^2+2leB}$ since the dominant contribution will come at the lowest Landau level and lowest $n$ we may go ahead and assume
\begin{equation}\label{eq:MQED_final}
M_\textrm{QED}^2  \sim\frac{e^3B}{\pi^{2}}\sqrt{\frac{\pi m_{e}}{2T}}e^{-\frac{m_{e}}{T}}\,.
\end{equation}

\subsection{ChPT Debye mass}

Having shown the Debye mass from the QED sector let us now turn to the pion sector. To determine the appropriate one-loop effective action we proceed along the lines outlined in ref.~\cite{Brauner:2021sci}, where thermal corrections to the CSL were explored. Namely, in a power counting scheme to $\mathcal{O}(p^{2})$ for the 1-loop correction, contributions from the WZW at $\mathcal{O}(p^{4})$ term may be neglected. Next, using the method outlined in the appendix of ref.~\cite{Brauner:2016pko}, we can determine the relevant fluctuation operators for both the neutral, $\mathfrak{D}^{(\pi^{0})}$, and charged pions, $\mathfrak{D}^{(\pi^{\pm})}$, whose respective effective actions read
\begin{equation}\label{eq:chptdet}
    \frac{i}{2}\mathrm{Tr}\ln(\mathfrak{D}^{(\pi^{0})})+i\mathrm{Tr}\ln(\mathfrak{D}^{(\pi^{\pm})})\,,
\end{equation}
c.f., eq.~\eqref{eq:qedeff} for the electromagnetic part. To determine the fluctuation operators we expand about the CSL of eq.~\eqref{eq:sineG} such that
\begin{equation}\label{eq:fluc}
    \Sigma=\Sigma_{0}U
\end{equation}
 where the fluctuations are captured with 
\begin{equation}
    U=\sqrt{1-\frac{\vec{\pi}^{2}}{f_{\pi}^{2}}}+\frac{i\vec{\tau}\cdot\vec{\pi}}{f_{\pi}}\,.
\end{equation}

One may be concerned that such an expansion is in conflict with the moduli approxi\-mation used to find the fermionic DWSks; consider eq.~\eqref{eq:moduli}. However, the moduli represent a zero-energy deformation about the family of degenerate classical solutions. And the fluctuations we seek here represent small oscillations about the classical solution and cost finite energy, in this case the CSL sine-Gordon solution. Further, the fluctuation we explore here also extend into the bulk in the $\hat{z}$ direction, whereas moduli of the CSL are independent of $z$; this is in fact our primary motivation. Nevertheless, for the temporal and spatial coordinates on the CSL plane the simultaneous treatment of moduli and fluctuations is acceptable so long as the moduli represent the slow collective motion about the CSL--this is in fact demanded in the Manton approximation, and the fluctuations represent fast modes.

Inserting eq.~\eqref{eq:fluc} into the ChPT Lagrangian given in eq.~\eqref{eq:LChPT} leads to a bilinear component for the pion fluctuations about the CSL given by~\cite{Brauner:2016pko}
\begin{equation}
\mathcal{L}_{\text{ChPT}}^{\text{bilin}}=\frac{1}{2}(\partial_{\mu}\vec{\pi})^{2}+(eA^{\mu}-\partial^{\mu}\chi^{3})(\pi_{1}\partial_{\mu}\pi_{2}-\pi_{2}\partial_{\mu}\pi_{1})+\frac{1}{2}e^{2}A^{2}(\pi_{1}^{2}+\pi_{2}^{2})-\frac{1}{2}m_{\pi}^{2}\vec{\pi}^{2}\cos\chi^{3}\,,
\end{equation}
where $A_{\chi}^{\mu}=eA^{\mu}-\partial^{\mu}\chi^{3}$. Let us next re-express the pion fields into their charged constituents so that $\pi^{\pm}=(1/\sqrt{2})(\pi_{1}\pm i\pi_{2})$. We will not need the neutral pion fluctuation operator for our analysis to follow. Then we can determine the charged pion part of bilinear part of the Lagrangian up to boundary terms as
\begin{align}
\mathcal{L}_{\text{ChPT}}^{\text{bilin}} & \subset\partial_{\mu}\pi^{+}\partial^{\mu}\pi^{-}+i(A^{\mu}-\partial^{\mu}\chi^{3})(\pi^{+}\partial_{\mu}\pi^{-}-\pi^{-}\partial_{\mu}\pi^{+})+(A^{2}-m_{\pi}^{2}\cos\chi^{3})\pi^{+}\pi^{-}\notag\\
 & =\pi^{+}\Bigl\{-(\partial-iA_{\chi})^{2}+(\partial_z\chi^{3})^{2}-m_{\pi}^{2}\cos\chi^{3}\Bigr\}\pi^{-}\,.
\end{align}
Then since the CSL dependent part of covariant derivative, $A_{\chi}^{\mu}$, is a total derivative, we can gauge away such a contribution in the effective action to find
\begin{equation}
\mathfrak{D}^{(\pi^{\pm})}=-(\partial-ieA)^{2}+(\partial_z\chi^{3})^{2}-m_{\pi}^{2}\cos\chi^{3}\,.
\end{equation}
The eigenspectrum we can see decouples into orthogonal and parallel to the magnetic field components, the former depending on the magnetic field, giving rise to a Landau level spectrum, and the latter characterized by the CSL.

Let us first determine the eigenspectrum, since the effective action may be expressed as a sum over all modes. The operator decomposes as $\mathfrak{D}^{(\pi^{\pm})}=-\partial_{0}^{2}+\mathfrak{D}_{B}^{(\pi^{\pm})}+\mathfrak{D}_{z}^{(\pi^{\pm})}$ with 
\begin{equation}
    \mathfrak{D}_{B}^{(\pi^{\pm})}=(\partial_{1}-ieA_{1})^{2}+(\partial_{2}-ieA_{2})^{2}\,.
\end{equation}
Its eigenvalues are well-established as the Landau levels for a scalar field
\begin{equation}
\lambda_{l}=(2l+1)eB\quad\forall l\in\mathbb{Z}^{+}\,.
\end{equation}
We next turn to the evaluation of the spectrum along the direction of the CSL in $z$. The operator here reads
\begin{equation}
\mathfrak{D}_{z}^{(\pi^{\pm})}=\frac{m_{\pi}^{2}}{\kappa^{2}}\Bigl[\partial_{u}^{2}-6\kappa^{2}\mathrm{sn}^{2}(u,\kappa)+\kappa^{2}+4\Bigr]\,,\label{eq:charged_pion_op}
\end{equation}
where we have made use of the coordinate $u=m_{\pi}z/\kappa$. The operator has an exact spectrum as a real Lame operator with $m=2$ gaps~\cite{lame2017}, which has five band edges. We refer to ref.~\cite{lame2017} for details on the derivation of the eigenspectrum of the real Lame operator. However, at this point, however, let us confine our attention to the kink limit, $\kappa\to1$, of the CSL for the calculation of the charged pion Debye length. In such a limit the charged pion operator in eq.~\eqref{eq:charged_pion_op} reduces to that of a P\"oschl-Teller potential with $\mathfrak{D}_{z}^{(\pi^{\pm})}=m_{\pi}^{2}[\partial_{u}^{2}+6\mathrm{sech}^{2}u-1]$, allowing for ultimately a tractable finite temperature expression. Let us use, however, the kink limit to be representative for the Debye length found in the ChPT sector; we will discover that the combined finite temperature screening is ultimately dominated through the QED sector.

Our goal is to first compute the zero temperature effective action defined for the scalar d.o.f. as
\begin{equation}
\mathcal{L}_{\text{ChPT}}^{1} =-i\int_{0}^{\infty}\frac{ds}{s}\langle x|e^{i(\mathfrak{D}^{(\pi^{\pm})}+i0^+)s}|x\rangle\,,
\end{equation}
from eq.~\eqref{eq:chptdet} for just the charged species as argued earlier, where we have also introduced an imaginary prescription for the causal operator that also gives us convergence in the infrared. Since the operator of the spacetime kernel is additive we may treat each portion separately, the magnetic part as a sum over Landau level, the energy as a timefree integral, and the P\"oschl-Teller part in the $z$ direction. The P\"oschl-Teller heat kernel fortunately is exactly known~\cite{Li:2015kha}. In the single soliton limit the eigenspectrum takes on a simple form with one zero mode, one bound state, a free part, and a continuum in the $\lambda_{i}$. The zero mode and bound state correspond to $\lambda_{i}=0,3m_{\pi}^{2}$. The free part follows as $L/\sqrt{4i\pi s}$ for characteristic length in the $z$ direction, $L$~\cite{Vassilevich:2003xt}. And finally there is a continuum part of the kernel characterized by a total phase shift $\delta(k)$ whose contribution enters as $\pi^{-1}\int^\infty_0 dk(d\delta(k)/dk)\exp(im_\pi^2(k^2+1)s)$. For the P\"oschl-Teller potential the total phase shift is given as $\delta(k)=2\arctan(2/k)+2\arctan(1/k).$~\cite{Li:2015kha}. Gathering each contribution we can write for the traced kernel in $z$ the following:
\begin{equation}
    K_z\coloneqq\int dz\langle z|e^{i\mathfrak{D}^{(\pi^{\pm})}_zs}|z\rangle=\frac{Le^{-im_\pi^2s}}{\sqrt{4i\pi s}}+1+e^{3im_{\pi}^{2}s}-\frac{1}{\pi}\int_{-\infty}^{\infty}dk\Bigl(\frac{1}{1+k^{2}}+\frac{2}{4+k^{2}}\Bigr)e^{-im_{\pi}^{2}(1+k^{2})s}\,.
\end{equation}
Then we may sum over the Landau levels and integrate over the energy to eventually find for the one-loop zero temperature effective action 
\begin{equation}
\mathcal{L}_{\text{ChPT}}^{1}  =-\frac{eB}{2\pi}\int_{0}^{\infty}\frac{ds}{s}\frac{\csc(eBs)}{\sqrt{-4\pi is}}\frac{K_z}{L}\,.
\end{equation}

Since the introduction of the operator component along $z$ introduces no new modification to the setup as shown for QED Debye length, we can write down the finite temperature 1-loop expression immediately. The only difference stems from the periodic BCs as opposed to the anti-periodic BCs, and thus here no $(-1)^n$ factor is present. We find for the one-loop thermal effective action the following:
\begin{equation}
\mathcal{L}_{\text{ChPT}}^{1T}=-\frac{eB}{2\pi}\int_{0}^{\infty}\frac{ds}{s}\frac{\csc(eBs)}{\sqrt{-4\pi is}}\frac{K_z}{L}\sum_{n=1}^{\infty}e^{i\frac{n^{2}}{4T^{2}}\frac{1}{s}}\cosh\Bigl(\frac{e\mu n}{T}\Bigr)\,.
\end{equation}
And the corresponding Debye length again expanding to the quadratic order in $\mu$, c.f. Eq.~\eqref{eq:m_Dformula}, becomes here
\begin{equation}
M_\textrm{ChPT}^2=-\frac{eB}{2\pi}\int_{0}^{\infty}\frac{ds}{s}\frac{\csc(eBs)}{\sqrt{-4\pi is}}
\frac{K_z}{L}\sum_{n=1}^{\infty}e^{i\frac{n^{2}}{4T^{2}}\frac{1}{s}}\Bigl(\frac{en}{T}\Bigr)^{2}\,.
\end{equation}

We can evaluate this expression as before for the QED case by summing over the Landau levels and then evaluating the propertime integrals in terms of modified Bessel functions.
\begin{align}
M_\textrm{ChPT}^2 &=\frac{e^3B}{\pi^2}\sum_{l=0}^\infty\sum_{n=1}^\infty
\frac{n}{T}\biggl\{\sqrt{m_\pi^2+(2l+1)eB}K_1\Bigl(\frac{n}{T}\sqrt{m_\pi^2+(2l+1)eB}\Bigr) \notag\\
 &\quad+\frac{\pi}{L}\Bigl[ e^{-\frac{n}{T}\sqrt{(2l+1)eB}}+e^{-\frac{n}{T}\sqrt{(2l+1)eB-3m_{\pi}^{2}}}\notag\\
 &\quad-\frac{1}{\pi}\int_{-\infty}^{\infty}dk\Bigl(\frac{1}{1+k^{2}}+\frac{2}{4+k^{2}}\Bigr)e^{-\frac{n}{T}\sqrt{(2l+1)eB+m_{\pi}^{2}(1+k^{2})}}\Bigr]\biggr\}\,.
\end{align}
Next, we can close the $k$ contour in the lower half imaginary plane, and evaluate the integral by taking the residue. We find due to the P\"oschl-Teller reflectionless property, the contribution from the discrete part is the exactly minus as that of the continuum and thus the Debye mass reduces to one of a scalar field in a magnetic field.
\begin{align}\label{eq:MChPT_full}
M_\textrm{ChPT}^2 &=\frac{e^3B}{\pi^2}\sum_{l=0}^\infty\sum_{n=1}^\infty
\frac{n}{T}\sqrt{m_\pi^2+(2l+1)eB}K_1\Bigl(\frac{n}{T}\sqrt{m_\pi^2+(2l+1)eB}\Bigr)\,. 
\end{align}
This is simply the quadratic Debye mass for a scalar field in a magnetic field to one-loop as anticipated. As a consistency check we may analyze the $B\to 0$  and high $T$ limit to find that $M_\textrm{ChPT}^2\sim e^2T^2/3$, exactly the same as for the spinor case, as is known. Eq.~\eqref{eq:MChPT_full} is exact to one-loop, however, as argued earlier in the QED case, we are chiefly concerned with the low temperature limit, and therefore let us look at the leading terms in such an expansion. Hence in the $T/\sqrt{m_\pi^2+eB}\ll 1$ limit such that only the lowest Matsubara number and lowest Landau level are treated we find
\begin{equation}
    M_\textrm{ChPT}^2\sim\frac{e^3B}{\pi^2}\sqrt{\frac{\pi\sqrt{m_\pi^2+eB}}{2T}}e^{-\frac{\sqrt{m_\pi^2+eB}}{T}}\,.
\end{equation}
We see this expression is quantitatively the same as the case for QED, eq.~\eqref{eq:MQED_final}, with $m_e\to\sqrt{m_\pi^2+eB}$. What this tells us is the thermal Debye quadratic mass is dominated by the one coming from QED since both are exponentially suppressed in the low temperature limit.

\subsection{Total Debye mass and finite temperature screening}

As we are concerned with the screening of only the fermionic domain-wall Skyrmions let us restrict our attention to the case in the zero temperature limit where Skyrmions first appear at $\mu_{B}>\pi\mathcal{C}(\kappa)$. There is indeed a finite Debye mass at finite temperature as evidenced in the previous two sections, and such terms can introduce a polarization of the vacuum impacting a vacuum polarization current, but will not affect the physical DWSk, which in ChPT in a magnetic field are identified with the baryons~\cite{Copinger:2025rpo}. Thus we can determine the total Debye mass follows as the sum coming from each sector, namely:
\begin{equation}\label{eq:Mtot}
    M_{\textrm{tot}}^2=M_{\textrm{Sk}}^2+M_{\textrm{QED}}^2+M_{\textrm{ChPT}}^2\,.
\end{equation}
Incorporating the Debye mass and hence interactions into the DWSk system lends itself to the interpretation of a fermi liquid; however as $M_{\textrm{tot}}$ becomes large the Debye length goes to zero and interactions become negligible leading to the previous picture of a fermi gas.

To determine the total and self-consistent screening characteristics let us examine the parts of the ensuing effective full (3+1)-dimensional Lagrangian concerning the quadratic in $a_0$ screening and the DWSk sources; this is 
\begin{equation}\label{eq:La_0}
    -\frac{e}{2}a_0\mathcal{B}+\frac{1}{2}a_0^2 M_\textrm{tot}^2-\frac{1}{4}f_{\mu\nu}f^{\mu\nu}\,.
\end{equation}
The first term is the electromagnetic coupling to the fermionic DWSk source written in terms of eq.~\eqref{eq:beta}, before having integrated out the half period of the CSL leading to the topological Skyrmion density term, i.e., $\int^{\ell /2}_0dz\mathcal{B}=q$. The second term is the term encompassing all the Debye screening at finite baryon chemical potential and finite temperature. We have further assumed here a weakly inhomogeneous $a_0$ accompanying the total Debye mass--as was assumed in a derivative expansion of the one-loop thermal effective actions. The dynamical gauge kinetic term reads for $f_{\mu\nu}=\partial_\mu a_\nu -\partial_\nu a_\mu$. The above Lagrangian furnishes the static screening potential argued previously. Namely, the classical e.o.m. in $a_0$ yields for $[\nabla^2-M_\textrm{tot}^2]a_0=-e\mathcal{B}/2$
\begin{equation}\label{eq:a_0}
    a_0(\boldsymbol{x})=-\frac{e}{2}\int d^3x' G(\boldsymbol{x}-\boldsymbol{x}')\mathcal{B}(\boldsymbol{x}')\,.
\end{equation}
The Green function here is
\begin{equation}
    G(\boldsymbol{x}-\boldsymbol{x}')=\int \frac{d^3k}{(2\pi)^3} \,e^{-i\boldsymbol{k}\cdot(\boldsymbol{x}-\boldsymbol{x}')}\frac{-1}{\boldsymbol{k}^2+M_\textrm{tot}^2}\,,
\end{equation}
which in configuration space satisfies a Yukawa-like screened potential about the Debye mass, $G(r)=\exp(-M_\textrm{tot}r)/(4\pi r)$ that provides the anticipated screening. However, the screening may only act in the CSL plane. Then inserting the above classical solution for $a_0$, eq.~\eqref{eq:a_0}, into eq.~\eqref{eq:La_0}, and retaining only the parts affected by the $a_0$ substitution we find the following term in the effective (3+1)-dimensional classical Lagrangian:
\begin{equation}\label{eq:q2}
    \frac{e^2}{8}\mathcal{B}(\boldsymbol{x})\int d^3x' G(\boldsymbol{x}-\boldsymbol{x}')\mathcal{B}(\boldsymbol{x}')\,.
\end{equation}
Next, we treat the situation in which the intra-Skyrmion separation is assumed to be much large than the total Debye length, and hence the Debye mass is assumed large such that a derivative expansion of the Green function may be applied; this is also a Thomas-Fermi approximation. In order to meet the large intra-Skyrmion separation we can either assume a finite temperature or for the zero temperature case avoidance of the single soliton limit such that $\kappa <1$. One then has in the effective Lagrangian that
$G(\boldsymbol{x}-\boldsymbol{x}')\approx-\delta(\boldsymbol{x}-\boldsymbol{x}')/M_\textrm{tot}^2$.
In which case eq.~\eqref{eq:q2} then becomes, within the \textit{(2+1) dimensional} effective Lagrangian,
\begin{equation}
    -\frac{e^2}{8M_\textrm{tot}^2}\mathcal{B}^2(\boldsymbol{x})\,.
\end{equation}
At this point, let us consider the term entering into the effective (2+1)-dimensional domain-wall Lagrangian. Thus we integrate out along a half period to find the above becomes
\begin{equation}\label{eq:q2final}
    -\frac{e^2}{8M_\textrm{tot}^2}\int^{\ell /2}_0dz\,\mathcal{B}^2(\boldsymbol{x})=-\frac{e^2}{M_\textrm{tot}^2}m_\pi \mathcal{C}_\textrm{TF} q^2(x,y)\,,
\end{equation}
where $\mathcal{C}_\textrm{TF}=[3/(8\pi\kappa)]{}_2F_{1}( -1/2,5/2;5;\kappa^2 )$. 

\begin{figure}
\centering
\includegraphics[width=0.7\columnwidth]{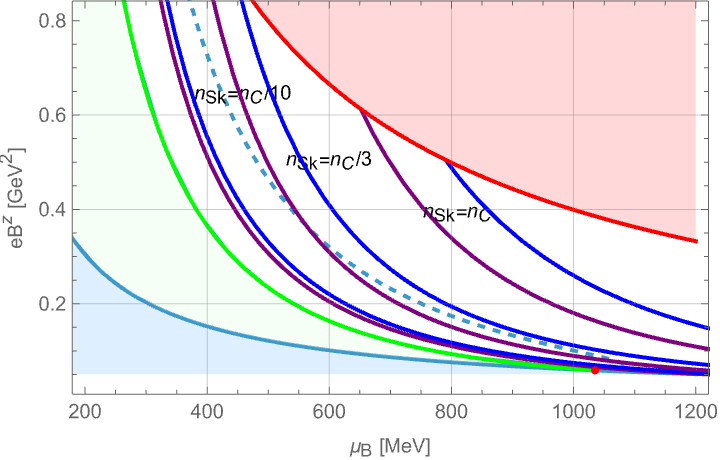}
\includegraphics[width=0.7\columnwidth]{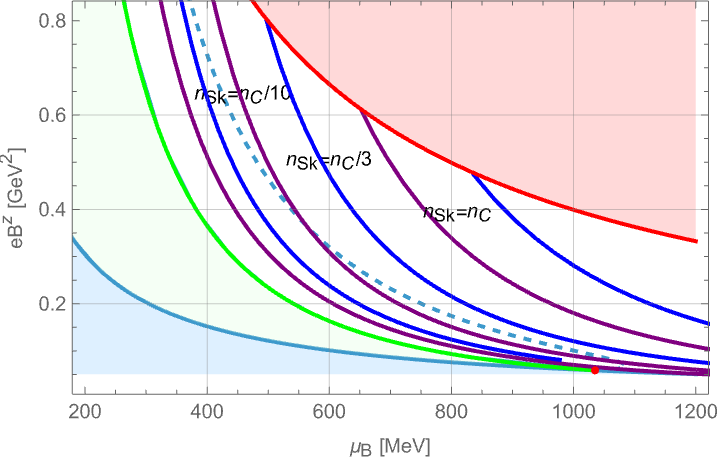}
\caption{Baryon density of the free quantum Fermi gas model, eq.~\eqref{eq:n_free}, and density including finite temperature screening, eq.~\eqref{eq:nSkT}, (top) and zero temperature screening (bottom) overlaid on top of the QCD phase diagram for finite chemical potential with a magnetic field. Densities are expressed in units of a nominal density $n_\textrm{C}=(\pi/m_{\pi})[8f_{\pi}^{2}/3]^{2}$. Isocurves of the free fermion gas model are shown as before in purple font, and the cases with Debye screening are shown in the blue font. For the finite temperature case (top) we have used for each screened isocurve a temperature of $T=m_\pi$. The green, red, light blue, and dotted light blue curves of the phase diagram are the same as in fig.~\ref{fig:phase}. Notice that for both zero and finite temperature cases isocurves lie increasingly to the right of those found with the free gas model. This shows that taking into account the effects of the screening through the various Debye masses given in eq.~\eqref{eq:Mtot} serve to shift the required baryon chemical potential needed to see the same density. Intuitively we can think about the addition of the shift of the required chemical potential for the zero temperature case as an introduction of a repulsive Coulomb interaction between the Skyrmions before being screened; this is what causes the increase of chemical potential for a fixed density. And for the finite temperature case since $M_\textrm{tot}$ increases the screening is increased and hence the isocurves more closely match the free fermi gas case. Theoretically if we could take the $M_\textrm{tot}\to\infty$ limit the free fermi gas case would be exact in the effective DWSk model. }
\label{fig:densityT}
\end{figure}

Now we are in a position to determine a self-consistent equation of state. Again as in sec.~\ref{sec:zero}, let us identify $n=-(\ell/2)q$ then we see eq.~\eqref{eq:q2final} acts as a shift of the baryon chemical potential by the density, giving us a self-consistent equation of state.
\begin{equation}\label{eq:mushift}
    \mu_B\to\bar{\mu}_B=\mu_B-\frac{e^2m_\pi\mathcal{C}_\textrm{TF}\ell}{2M_\textrm{tot}^2}n\,.
\end{equation}
Then, as we accomplished for the free gas case in eq.~\eqref{eq:nSk_def}, we can write down a free equation of state where $n=(\bar{\mu}_B^2-m_\textrm{Sk}^2)/(2\pi\ell)$ forms a self-consistent equation for the Skyrmion density
\begin{equation}
    n=\frac{1}{2\pi\ell}\Bigl\{ \Bigl( \mu_B-\frac{e^2m_\pi\mathcal{C}_\textrm{TF}\ell n}{2M_\textrm{tot}^2}\Bigr)^2 -m_\textrm{Sk}^2\Bigr\}\,.
\end{equation}
The above can be expanded into a quadratic equation in $n$, whose solution reads
\begin{equation}\label{eq:nSkT}
    n_{\textrm{Sk}}=\frac{1}{\ell}\frac{\mu_{B}^{2}-m_{\textrm{Sk}}^{2}}{\pi+\mu_{B}\alpha_{\textrm{TF}}+\sqrt{\pi^{2}+2\pi\mu_{B}\alpha_{\textrm{TF}}+m_{\textrm{Sk}}^{2}\alpha_{\textrm{TF}}^{2}}}\,,
\end{equation}
where $\alpha_\textrm{TF}=e^{2}m_{\pi}\mathcal{C}_{\textrm{TF}}/(2M_{\textrm{tot}}^{2})$. Because we can see that as the density increases, the electromagnetic screening will induce an effective density dependent interaction between the DWSks in the crossover to a fermi liquid. Let us see how the revised density compares to the one found before at zero temperature using a free fermion gas model in eq.~\eqref{eq:n_free}. Immediately we can tell that in the limit $\alpha_\textrm{TF}\to 0$ the density from the free gas model appears, which is consistent with a large total Debye mass. However, to get a better picture of the comparison between the two models let us plot both their isocurve densities at a similar density, both for the zero temperature and finite temperature cases. This is done in fig.~\ref{fig:densityT}.
The plot shows all solutions to the coupled set of equations for both the density in eqs.~\eqref{eq:n_free} and~\eqref{eq:nSkT}, and constraint between the magnetic field and baryon chemical potential given by eq.~\eqref{eq:kappa_constraint}.\footnote{This involves numerically evaluating the roots of the coupled equations; since $M^2_\textrm{ChPT}$ is suppressed by the pion mass and $M^2_\textrm{QED}$ by the electron mass, screening from the QED sector will suppress any screening from the ChPT sector. Therefore, to sensibly numerically evaluate for the roots we take a nominal value of chemical potential governing the magnetic field in the pion mass term in $M^2_\textrm{ChPT}$ as $\sqrt{m_{\pi}^{2}+eB}\to\sqrt{m_{\pi}^{2}+16\pi f_{\pi}^{2}m_{\pi}E(\kappa)/(\kappa 700 \text{MeV})}$; because of the suppression the density found from the coupled equations is insensitive to whatever value is used here.} Lines in purple as before represent the free fermion gas model, and the lines in blue are the density isocurves found from eq.~\eqref{eq:nSkT} for either the zero temperature case, the bottom figure, or the finite temperature case, the top figure. We have used a temperature here of $T=m_\pi$ for the finite temperature case. Notice that the free fermi gas model remains a good predictor of the density for lower values of the density for any temperature; see the curves at $n_\textrm{Sk}=n_C/10$. However, for increasing density the free gas model predicts a lower baryon chemical potential than does the full screening model. This makes sense according to eq.~\eqref{eq:mushift} where we see that the required chemical potential setting the Fermi wavenumber is reduced by the increased density induced screening. Qualitatively, however both models predict similarly shaped density isocurves. An intuitive way to think about the results here is to first consider the free gas model with $\mu\to\varepsilon_F$, for the Fermi energy associated with Fermi momentum given in eq.~\eqref{eq:kF}. Then the addition of the \textit{zero temperature} screening amounts to the introduction of a repulsive interaction between the DWSks such that $\mu\to \varepsilon_F+V$, for potential $V$. When we applied the Thomas-Fermi approximation of screening we can then see that to achieve the same density the chemical potential must be raised, this is even more so as the temperature raises, where the required chemical potential for a given density would decrease in comparison to the zero temperature case.

\section{Summary and discussion}
\label{sec:conclusions}

We have examined the screening of fermionic DWSk leading to expressions for anticipated densities of the DWSks, and hence baryon density, in regions of interest for the QCD phase diagram. Concretely we have studied two flavor ChPT, valid for low-energy QCD, coupled to QED and moreover under a baryonic chemical potential and strong magnetic field. The magnetic field brings about a QCD vacuum state of layered domain wall, the CSL, and by studying moduli around the CSL one discovers a DWSk phase in which baryons form on the half solitonic period of the CSL. As a first approximation, and at zero temperature, we study equilibrium densities of DWSk by treating the fermionic DWSk as a free fermionic quantum gas on the (2+1)-dimensional sheet of the CSL, in which all available quantum states are filled to (2+1)-dimensional Fermi energy on the disk. We have found eq.~\eqref{eq:n_free}, predicting a density entering at quadratic order in the baryonic chemical potential minus an effective DWSk mass found at the critical chemical potential characterizing the DWSk phase transition, or at $m_\textrm{Sk}=\mu_c$. 

To treat Debye screening, we have further incorporated QED, a dynamical photon, and the effects of finite temperature leading to a Debye screening from vacuum polarization, also at finite temperature. This is done using a derivative expansion of the full QED effective action; leading order corrections appear at a constant $a_0$, and inform us of the Debye length. A similar treatment has been performed on fluctuations about the CSL for the ChPT sector. And a similar Debye length in comparison to the QED case has been found. We additionally found a Debye length associated with the screening of DWSks among themselves at zero temperature. The effects of including the sum total of the quadratic Debye masses leading to a finite temperature density has been argued in eq.~\eqref{eq:nSkT}, where qualitatively similar behavior in comparison to the free fermion gas model at low density has been found.

Let us also remark that the zero-temperature Debye mass provides a quantitative criterion for the validity of the dilute Fermi-gas description. The corresponding Debye length is $\lambda_{\rm Sk}=M_{\rm Sk}^{-1}$, from which one may define a characteristic density $n_\lambda\sim\lambda_{\rm Sk}^{-3}$ beyond which screening effects become significant. Evaluating this estimate over the range of applicable CSL solutions as a function of $\kappa$, one finds that the corresponding isocurve lies entirely inside the red region of Fig.~\ref{fig:phase}, where ChPT is no longer applicable. Therefore, throughout the regime in which ChPT is reliable, the dilute Fermi-gas approximation employed in this work remains self-consistent.

Beyond the dilute regime, however, the physical picture is expected to change qual\-itatively. As the Skyrmion density increases, the mean inter-Skyrmion distance eventually becomes comparable to the Debye screening length, so that the screening clouds of neigh\-boring DWSks begin to overlap and the non-interacting Fermi-gas approximation breaks down. The system then enters a strongly interacting many-body regime. One natural possibility is a crossover to a Landau Fermi liquid, in which the quasiparticles remain well defined but acquire nontrivial effective interactions. Alternatively, if the screened Coulomb repulsion becomes sufficiently strong, the DWSks may instead favor spatial ordering and form a crystalline phase analogous to a Wigner crystal. Determining the effective interaction between DWSks and clarifying which of these scenarios is realized lies beyond the scope of the present work and remains an interesting direction for future investigation.

It is also intriguing to speculate that, in the strongly interacting regime, two fermionic DWSks may form a bound state corresponding to the bosonic DWSk previously identified in the effective theory over a full CSL period. If such bound states become energetically favorable, they could undergo Bose-Einstein condensation, providing yet another possible many-body phase beyond the dilute Fermi-gas regime. We leave this interesting possibility for future work.

\section*{Acknowledgments}

This work is supported in part by JSPS Grant-in-Aid for Scientific Research KAKENHI Grant No. JP22H01221 and JP23K22492 (M.~E., M.~N.). 
This work is also supported in part by the WPI program ``Sustainability with Knotted Chiral Meta Matter (WPI-SKCM$^2$)'' at Hiroshima University. P.C. would further like to acknowledge support from the Research Start-up Support Fund of WPI-SKCM$^2$.

\bibliographystyle{JHEP}
\bibliography{references}

\end{document}